\documentstyle[12pt]{article}
\begin{document}
\baselineskip24pt
\title{Cauchy Noise and Affiliated Stochastic  Processes}

\author{Piotr Garbaczewski\\
Institute of Theoretical Physics, University of Wroc{\l}aw\\
PL-50 204 Wroc{\l}aw, Poland\\
and\\
Robert Olkiewicz\thanks{permanent address: Institute of Theoretical Physics,
 University of Wroc{\l}aw, PL 50-204 Wroc{\l}aw, Poland}\\
Fakult\"{a}t f\"{u}r Physik, Universit\"{a}t Bielefeld\\
D-33 615 Bielefeld, Germany}
\maketitle

\begin{abstract}
By departing from the previous attempt (Phys. Rev. {\bf E 51},
4114, (1995))  we give a detailed construction of conditional and
perturbed Markov processes,  under the assumption that the Cauchy law
of probability replaces the Gaussian law (appropriate for the
Wiener process)  as the model of primordial noise.   All considered
processes  are regarded as probabilistic solutions of the so-called
Schr\"{o}dinger interpolation problem, whose validity is thus extended
to the jump-type processes and their step process approximants.

\end{abstract}


\section{Introduction}

Probabilistic solutions of the so-called Schr\"{o}dinger boundary data
problem, \cite{schr,jam},  are known to yield  a unique Markovian
interpolation between any two strictly positive probability densities
designed to form the input-output statistics data for a certain
dynamical process, taking place in a finite-time interval.
The key problem, if one attempts to  reconstruct  the \it most
likely \rm   (Markovian) dynamics,
is to select the jointly continuous in space  variables positive 
and contractive semigroup   kernel. That issue was analyzed before
in a number of publications, \cite{schr}--\cite{blanch}.

In fact, basic stochastic processes of the nonequilibrium statistical
physics (Smoluchowski diffusion processes) involve    the familiar
Feyman-Kac-like kernels as the building blocks for  suitable Markovian 
transition probability densities, \cite{zambr,olk,kondrat}.
In the standard "free" case (Feynman-Kac potential
equal to zero as a necessary condition) the  Wiener noise
may  be  recovered.

In the framework of the Schr\"{o}dinger problem the choice of the
integral  kernel  is arbitrary,
except for the strict positivity (cf. however, \cite{blanch})
and continuity demand.
It is thus rather natural to ask for the most general
stochastic interpolation, that is admitted under the above premises.

Clearly, the standard Feynman-Kac
kernels generated by Laplacians plus suitable potentials, \cite{sim,simon},
  are very special examples in
a surprisingly rich encompassing family. First of all, the
concept of the Gaussian noise,  regarded as a stochastic analogue of
the mechanical "state of rest" and  traditionally linked with  a Wiener process,
can be extended to
all infinitely divisible probability laws via the L\'{e}vy-Khintchine formula.
It expands our framework from continuous diffusion processes to  jump 
or combined diffusion--jump propagation scenarios, \cite{klaud}, as appropriate
mathematical models of the primordial "free noise".

The next natural step in the analysis is to account for typical
perturbations of any given  process, according to  the   pattern of the
Feynman-Kac formula hence in terms of perturbed semigroups, where
an appropriate   generator (replacing the
Laplacian) is additively modified by a suitable potential.

The  Feynman-Kac formula
is known to extend its validity to the pertinent non-Gaussian
 measures, \cite{carm,carm1,jacob}.
However, to our knowledge, \it no \rm  detailed description of the related
Feynman-Kac \it kernels \rm , with emphasis on their continuity
and positivity (those features must be settled, \cite{jam,zambr,olk},
in the context  of
the above-mentioned Schr\"{o}dinger interpolation problem), exists
in the literature.
Quite in contrast  with the elaborate analysis that is available 
with respect to the  Wiener measure, \cite{simon}.

By referring to a physical terminology, let us consider  Hamiltonians
(semigroup generators)
of the form $H=F(\hat{p})$, where
$\hat{p}=-i
\nabla $ stands for the momentum operator and  for 
$-\infty <k<+\infty $, $F=F(k)$ is a real valued, 
bounded from below, locally integrable function. Here, $ \hbar = c= 1$.
We simplify  further  discussion by considering  processes in one
spatial dimension. We easily learn that for times $t\geq 0$ there holds
$${[exp(-tH)]f(x)
 = [exp(-tF(p)) \hat{f}(p)]^{\vee }(x)}\eqno (1)$$
where the superscript $\vee $ denotes the inverse Fourier transform and
  $\hat{f}$  stands for the Fourier transform of $f$.
 
Let us set $k_t={1\over {\sqrt {2\pi }}}[exp(-tF(p)]^{\vee }$, then the 
action of $exp(-tH)$ can be given in terms of a convolution: 
$exp(-tH)f = f*k_t$, where $(f*g)(x): =\int_R g(x-z)f(z)dz $. 

 We are interested in those $F(p)$ which give rise to
 positivity preserving semigroups:  if $F(p)$ satisfies the
 L\'{e}vy-Khintchine formula, then $k_t$ is a positive measure for all 
 $t\geq 0$.
Let us concentrate on the integral part of the L\'{e}vy-Khintchine formula,
which is responsible for arbitrary stochastic jump features:
$${F(p) = -  \int_{-\infty }^{+\infty } [exp(ipy) - 1 - 
{ipy\over {1+y^2}}] 
\nu (dy)}\eqno (2)$$
where $\nu (dy)$ stands for the so-called L\'{e}vy measure.

There are not many explicit examples (analytic formulas for probability
densities) for processes governed  by (2), except possibly for
the so called stable probability laws. The best known example is the
classic Cauchy density.
Let us focus our  attention on that selected choice  for the
characteristic exponent $F(p)$, namely: 
$F_0(p)=|p|$ which is the Cauchy process generator.
The  semigroup generator  $H_0$ is a
pseudodifferential operator. The associated kernel $k_t$
in view 
of the "free noise" restriction (no potentials at the moment)
is  a transition density of
the jump-type (L\'{e}vy) process, determined by
 the  corresponding  L\'{e}vy measure $\nu (dy)= {1\over \pi }
 {dy\over {y^2}}$.

It is instructive to notice that a pseudodifferential analog of the
Fokker-Planck equation holds true:
${F_0(p)\Longrightarrow \partial _t{\overline {\rho }}(x,t)=
 - |\nabla|{\overline {\rho }}(x,t)}$.
 This evolution rule gives rise to
 the Cauchy process  probability density  $\rho
 (x,t)={1\over \pi }{t\over {t^2+x^2}}$ and the corresponding
 space-time homogeneous transition  density (e.g.
 the semigroup kernel in this free propagation case).

As mentioned before,  the existence and uniqueness of
 solutions proof for the Schr\"{o}dinger problem  extends, \cite{klaud},
  to  cases governed by  infinitely divisible probability laws.

Our principal
goal in the present paper is to generalise this observation 
  to encompass the additive perturbations by  physically motivated 
 potentials. The construction is  based on the
 Feynman-Kac formula for perturbed semigroups, with strictly
 positive and jointly continuous kernel functions.

As a byproduct of the discussion we shall give a
characterisation of the affiliated Markovian jump-type processes
in terms of approximating  (convergent)  families of
more traditional, \it step \rm processes, that solve a suitable version
of the Schr\"{o}dinger interpolation problem.

The demonstration explicitly
pertains to the  Cauchy process and its relatives, albeit the 
techniques and major statements may be extended to a broader  class of
L\'{e}vy  processes and their perturbed versions, cf.
\cite{klaud,carm,carm1} and \cite{shl}-\cite{grigo} for related mathematical
and physical connotations.

\section{The Cauchy process and its conditional relatives}

We  consider Markovian propagation scenarios so remaining within
the well established framework, where the input-output statistics data
are provided in terms of two  strictly positive boundary
 densities $\rho (x,0)$
and $\rho (x,T)$, $T>0$. In addition, a  bi-variate transition probability
density is given in a specific factorized
form:  $m(x,y)=f(x)k(x,0,y,T)g(y)$, with marginals:
$${\int_R  m(x,y) dy =\rho (x,0)\, ,\, \int_R m(x,y)dx =\rho (y,T)}
\eqno (3)$$
Here, $f(x), g(y)$ are the a priori unknown functions, to come out as
strictly positive solutions of the integral  system of equations (3),
provided that in addition to the density boundary data we have in
hands  {\it any}  strictly positive, jointly continuous  in space
variables {\it function }  $k(x,0,y,T)$.
Additionally, we impose a restriction that $k(x,0,y,T)$
 represents  a  certain strongly
continuous dynamical semigroup  kernel $k(y,s,x,t), 0\leq s\leq t<T$ ,
while given at the time interval borders:  it  secures  the
Markov property of the sought for stochastic process.

Under those circumstances, \cite{olk}, once we  define functions
$${\theta (x,t)=\int dy k(x,t,y,T)g(y)}, 
\qquad\theta _*(y,s)=\int dx k(x,0,y,s)f(x) \eqno (4)$$
there exists  a transition density
$${p(y,s,x,t)=k(y,s,x,t){{\theta (x,t)}\over {\theta (y,s)}}}
\eqno (5)$$
which implements a  Markovian propagation of the
probability density
$${\rho (x,t)= \allowbreak
\theta (x,t)\allowbreak \theta _*(x,t)}\eqno (6)$$
$${\rho (x,t) = \int p(y,s,x,t)\rho (y,s)dy} $$
between the prescribed boundary data.

For a given semigroup which is characterized by its generator
 (Hamiltonian), the kernel $k(y,s,x,t)$ and the emerging transition
 probability density
$p(y,s,x,t)$  are unique in view of the uniqueness of solutions
$f(x),g(y)$. For Markov processes, the knowledge of the
transition probability density $p(y,s,x,t)$ for all intermediate
times $0\leq s< t\leq T$  suffices
for the derivation of all other relevant characteristics. 

At this point, let us make a definite choice of the kernel function,
namely that of  the Cauchy kernel:
$${k(y,s,x,t)={1\over {\pi }} {{t-s}\over {(t-s)^2 + (x-y)^2}}\, .}
\eqno (7)$$

We have:\\

{\bf Theorem 1}: \\
(a) $p(y,s,x,t)$ defined by Eqs. (5) and (7)  is a Markov transition
kernel, that is (weak limit in below)\\
$$\int_R p(y,s,x,t) dx = 1$$
$$lim_{t\downarrow s} p(y,s,x,t) = \delta _y(x)$$
$$ \int_R p(y, t_1,z,t_2)p(z,t_2,x,t_3)dz = p(y,t_1,x,t_3) $$
for all $0\leq t_1<t_2<t_3\leq T$, with $\delta _y$ standing for the
Dirac delta \\
(b) $\rho (x,t)$, Eq. (6), is a probability distribution interpolating
between $\rho _0$  and $\rho _T$:\\
$$\int_R \rho (x,t) dx = 1$$
$$\rho (x,0)= \rho _0(x),\, \rho (x,T)=\rho _T(x)$$
(c) the process $X_t$ having $p(y,s,x,t)$ as the transition kernel is
a Markov  interpolating process:
$$\int_R p(y,s,x,t)\rho (y,s) dy = \rho (x,t)$$
for all $0\leq s< t \leq T$.\\

{\bf Proof}:   See e.g. Refs. \cite{klaud,olk}.   \\

 Let us notice that the process $X_t$ is obtained from the
Cauchy process $X^C_t$ by means of a multiplicative transformation of
transition function. Clearly,  $\alpha ^s_t= {{\theta (X^C_t,t)}\over
{\theta (X^C_s,s)}}$ is a multiplicative functional of $X^C$ such that
its average with respect to the Cauchy process reads
 $\int \alpha ^s_t(\omega ) P^C_x(d\omega )=1$ for any
 $0\leq s\leq t\leq T$ and any $x\in R$, see e.g.
\cite{dynkin}.  However $\alpha ^s_t$ is not homogeneous and, even worse,
not  contracting (in fact, not even  bounded).
We cannot be a priori sure that the generic sample path properties of the
Cauchy process  can be attributed to $X_t$ as well.
In particular, an approximation of $X_t$ in terms of jump processes with a
finite number of jumps in a finite time interval, is by no means
obvious and needs a  demonstration (to be given in below).

To this end, let us first notice that $\theta _*$ and $\theta $ satisfy
the conjugate pseudodifferential equations:
$$\partial _t \theta _* = -|\nabla | \theta _* $$
$${\partial _t\theta =|\nabla | \theta }\eqno (8)$$
where the operator $|\nabla |$ acts as follows:
$${|\nabla |f(x) = - {1\over \pi }\int_R [f(x+y) - f(x) -
{{y\nabla f(x)}\over {1+y^2}}] {dy\over y^2}\, .}\eqno (9)$$

Let us define a new operator $|\nabla |_\epsilon $ by:
$${|\nabla |_\epsilon f(x)= - {1\over \pi } \int_{|y|>\epsilon }
[f(x+y) - f(x)]{dy\over y^2}}\eqno (10)$$
and, accordingly:
$$\partial _t \theta _*^\epsilon = - |\nabla |_\epsilon \theta _*^\epsilon $$
$${\partial _t \theta ^\epsilon = |\nabla |_\epsilon
\theta ^\epsilon }\eqno (11)$$
with $\theta _*^\epsilon (x,0)=\theta _*(x,0)$,
$\theta ^\epsilon (x,T)=\theta (x,T)$.

Furthermore, let
$${q_\epsilon (x) = {1\over \pi }
\chi _{I_\epsilon ^c}(x) {1\over x^2}}\eqno (12)$$
where  $I_\epsilon ^c = [-\epsilon ,
\epsilon]^c = \{ x\in R: |x|> \epsilon \}$ and $\chi _A$ is an
indicator function of a set $A$.\\

We have: \\

{\bf Theorem 2}:\\
Let us define the Poisson transition kernel corresponding to
the measure $q_\epsilon (x)dx$:\\
$$k_\epsilon (x,t)= [exp(-{2t\over {\epsilon \pi }})]\, [\delta _0(x) + tq_\epsilon (x) +
{t^2\over 2!}(q_\epsilon * q_\epsilon )(x) + ...]\, .$$
 Then, functions: \\
$$\theta _*^\epsilon (x,t) = \int_R k_\epsilon (x-y,t)\theta _*(y,0)dy$$
$$\theta ^\epsilon (x,t) = \int_R k_\epsilon (x-y, T-t)\theta (y,T)dy$$
solve the Cauchy  problem (11).\\

{\bf Proof}: \\

The transition function in the above is called the Poisson transition kernel
following the terminology of Ref. \cite{dynkin}. 
We have $\theta _*^\epsilon (x,0) = \int_R \delta _0(x-y) \theta _*(y,0)dy =
\theta _*(x,0)$ and:
$$\partial _t\theta _*^\epsilon (x,t) = \int _R [\partial _t k_\epsilon (x-y,t)]
\theta _*(y,0) dy $$
where
$$\partial _t k_\epsilon (x,t) = - {2\over {\pi \epsilon }} k_\epsilon (x,t) +
[exp(-{2t\over \epsilon })] \, [q_\epsilon (x) +
 t (q_\epsilon * q_\epsilon )(x) + ...]\, .$$
 Consequently,
$$[\partial _t k_\epsilon (.,t) * \theta _*(.,0)](x) = -{2\over {\pi \epsilon }}
\theta _*^\epsilon  (x,t) + [[exp(-{2t\over \epsilon })]
q_\epsilon * (\delta _0 + tq_\epsilon +
{t^2\over {2!}}q_\epsilon *q_\epsilon +...)*\theta _*](x) = $$
$$-{2\over {\pi \epsilon }} \theta _*^\epsilon  (x,t) +
[ q_\epsilon * \theta _*^\epsilon (.,t)](x)=
-{2\over {\pi \epsilon }} \theta _*^\epsilon (x,t) +
\int _R q_\epsilon (y)
\theta _*^\epsilon  (x-y, t)dy \, .$$
But, there holds
$$\int_R q_\epsilon (y) \theta _*^\epsilon  (x-y,t) dy = {1\over \pi }
\int _{|y|>\epsilon }
 \theta _*^\epsilon ((x-y,t) {dy\over y^2} =
{1\over \pi }\int_{|y|>\epsilon } \theta _*^\epsilon  (x+y,t)
{dy\over y^2}$$
and, in view of the obvious identity
$${{2\over {\pi \epsilon }}\theta _*^{\epsilon }(x,t) =
{1\over \pi } \int_{|y|>\epsilon }
\theta _*^{\epsilon }(x,t) {dy\over y^2}}\, ,$$
 we finally arrive at
$$\partial _t\theta _*^{\epsilon }(x,t) = {1\over \pi } \int_{|y|>\epsilon }
[\theta _*^\epsilon (x+y,t) - \theta _*^\epsilon (x,t)] {dy\over y^2} =
- |\nabla |_\epsilon \theta _*^{\epsilon }(x,t)$$
An analogous line of arguments follows with respect to
$\theta ^\epsilon (x,t)$,  which completes the proof.\\

A random process  with a Poisson transition function belongs to the
class of, so called,  step processes,
\cite{dynkin,gihman}, that is   jump processes with no accumulation points
of jumps in a finite time interval: the number of jumps
is finite on each finite time interval.
We have: \\

{\bf Lemma 1}: \\
The Markov process $Y_t^\epsilon $ given by the transition function
$k_\epsilon (x,t)$  is a step process with  a characteristic function:
$$\Phi _\epsilon (p,t) = exp (-t [\hat{q}_\epsilon (0) -
\hat{q}_\epsilon (p)])$$
where $\hat{q}_\epsilon (p) $ is the Fourier transform of $q_\epsilon (x)$.\\

{\bf Proof}:\\

We need to evaluate the characteristic function of the transition kernel,
that is :
$$\Phi _\epsilon (p,t) = exp(-{{2t}\over {\pi \epsilon  }})\cdot
\int_{-\infty }^{+\infty } [exp(-ipx)]  \, [\delta _0(x) + tq_\epsilon (x) +
{t^2\over {2!}}(q_\epsilon *q_\epsilon )(x) +...] dx =$$
$$exp(-{{2t}\over {\pi \epsilon }})\cdot  [1+ t\hat{q}_\epsilon (p) +
{t^2\over {2!}}(\hat{q}(p))^2+ ...]= exp[-{2t\over {\pi \epsilon }}+
t\hat{q}_\epsilon (p)]$$
In view of  $\hat{q}_\epsilon (0) = {2\over {\pi \epsilon }}$,
the Lemma holds true.      \\

As a technical warming up we shall now prove that the Cauchy process
is the limit ( in distributions)   of a one-parameter family of step
processes $Y_t^\epsilon $. We touch here an important issue  of
limits (convergence) of jump processes, \cite{brei,bil,jacod} and
there are many types of the pertinent convergence.  For example,
 it is known
that $Y_t^\epsilon $   tends to the Cauchy process  in probability,
\cite{brei}, while  major modern techniques  refer to
 the   weak convergence of probability measures,
\cite{jacod}). Also, typical proofs refer only to processes with stationary
 independent increments,  while  we cannot respect this limitation
 in the presence of  perturbations.\\

{\bf Lemma 2}:\\
There holds: $lim_{\epsilon \rightarrow 0} \Phi _\epsilon (p,t) = \psi (p,t)$,
where $\psi (p,t)$ is the Cauchy  characteristic function $\psi (p,t) = exp(-t|p|)$.
Moreover, the limit is uniform for all $t\in [0,T]$.\\

{\bf Proof}: \\

 Let us evaluate $\hat{q}_\epsilon (p)$:
$$\hat{q}_\epsilon(p) = {1\over \pi }\int_{-\infty }^{+\infty } exp(ipx) \cdot
\hat{q}_\epsilon (x) dx = {1\over \pi }\int_{|x|>\epsilon }
exp(ipx)\cdot {dx\over x^2} = $$
$${2\over \pi } \int_{\epsilon }^{\infty } {{cos(px) - 1}\over x^2}dx +
{2\over {\pi \epsilon }}$$
Consequently
$$\Phi _\epsilon (p,t)= exp[ - {2t\over \pi } \int_{\epsilon }^{\infty }
{{1-cos(px)}\over x^2} dx]\, .$$
In view of
$$lim_{\epsilon \rightarrow 0} \int_{\epsilon }^{\infty }
{{1-cos(px)}\over x^2}dx = {{|p|\pi }\over 2}\, ,$$
we arrive at:
$$lim_{\epsilon \rightarrow 0} \Phi _{\epsilon }(p,t) = exp [- {2t\over \pi } \cdot
{|p|\pi \over 2}]=exp(-t|p|)\, .$$
The proof is completed. \\

Clearly, $|\nabla |_\epsilon$ is a well defined
semigroup generator for the step process $Y_t^\epsilon $. Let us recall
that  sample paths of a step process have only a finite number of jumps in each
finite time interval, and between jumps the sample path is constant,
\cite{gihman}.
The limiting Cauchy process   belongs to the category of jump-type processes,
where apart from  the long jumps-tail (no fixed bound can be imposed on their
length) that implies the nonexistence of
moments of the probability measure, sample paths of the Cauchy process
may have an infinite number of jumps of arbitrarily small size.
By general arguments,
pertaining to the space $D_E[0,\infty )$ of right continuous functions with
left limits (cadlag),  both in the finite and ifinite time interval
the number of jumps is at most countable, \cite{brei,kurtz}.
It is also useful to recall that on a finite time interval there can be
at most finitely many points $t\in [0,T]$ at which the jump size exceeds
a given positive number. In view of that,  $sup_{t\in [0,T]}\,
|Y_t^\epsilon | < \infty $.  Obviously,  there is no fixed upper  bound
for the size of jumps (except for being finite), since  a stochastically
continuous process with independent increments having, with probability 1,
no jumps exceeding a certain constant $C$, would possess all moments,
\cite{gihman}.

Now, we shall pass to a slightly more involved demonstration that a well
defined family of Markov processes $X_t^\epsilon $ (in fact, step ones) can be
constructed,  such that the process $X_t$ of Theorem 1 can be approximated
(in the sense of suitable convergence) to an arbitrary degree of accuracy.
\\
Here, we are motivated by a heuristic analysis carried out in our
 earlier paper, \cite{klaud}. There,  we have found that after  neglecting
 "small jumps", the time
evolution of the resultant probability density $\bar{\rho }_\epsilon $
may be written as:
$${\partial _t \bar{\rho }_\epsilon (A,t) = \int_R q_\epsilon (t,x,A)
\bar{\rho } _\epsilon (x,t) dx + <v>_A(t) \int_{|y|>\epsilon }
{y\over {1+y^2}}d\nu (y)\, .}\eqno (13)$$
The measure $d\nu $ is symmetric around the point $\{0\}$, hence the second
term cancels, and we arrive   at
$${\partial _t \bar{\rho }_\epsilon (A,t) = \int_R q_\epsilon (t,x,A)
\bar{\rho } _\epsilon (x,t) dx}\eqno (14)$$
where   the so-called jump intensity reads
$${q_\epsilon (t,y,A)= \int_{|y|>\epsilon} {{\theta ^\epsilon (y+x,t)}\over
{\theta ^\epsilon (y,t)}}[ \chi _A(x+y) - \chi _A(y)]d\nu (x)}\eqno (15)$$
and $\theta ^\epsilon (x,t)$ comes out as a solution of the
second pseudodifferential equation in the formula (11).  \\

Let us define (cf. Eq. (12))
$${h_\epsilon (t,y)= \int_{-\infty }^{+\infty } {{\theta ^{\epsilon }
(x+y,t)}\over
{\theta ^{\epsilon }(y,t)}} q_\epsilon (x) dx}\eqno (16)$$
and
$${h_\epsilon (t,y,x) = {{\theta  ^{\epsilon }(x,t)}\over
{\theta ^{\epsilon }(y,t)}} q_\epsilon (x-y)\, .}\eqno (17)$$
Then, clearly   the jump intensity (14) takes the form
$${q_\epsilon (t,y,A) = \int_A h_\epsilon (t,y,x)dx -
h_\epsilon (t,y) \chi _A(y)} \eqno (18)$$

With those notations, we have: \\

{\bf Lemma 3} \\

If the function   $g(y)$ (cf. Eq. (4)) is uniformly bounded, then
$h_\epsilon (t,y,x)$ is a density of a finite measure and $h_\epsilon (t,y) =
\int _R h_\epsilon (t,y,x) dx$.\\

{\bf Proof}: \\

By our assumption,  $g(y)\leq M$ for all $y\in R$.
Because of $\theta ^\epsilon
(x,t) = \int _R k_\epsilon (T-t,x-y) g(y) dy $, we have a bound
$$\theta ^\epsilon (x,t) \leq M\, \int_R k_\epsilon (T-t,x-y) dy = M\, .$$
Hence
$$\int_{-\infty }^{+\infty } h_\epsilon (t,y,x)dx = \int_{-\infty }^{+\infty }
{{\theta ^\epsilon (x,t)}\over {\theta ^\epsilon  (y,t)}}
q_\epsilon (x-y)dx= h_\epsilon (t,y)$$
and
$$h_\epsilon (t,y) \leq  {M\over {\theta ^\epsilon (y,t) }}
{2\over \epsilon }\, .$$
It is also clear that $h_\epsilon (t,y,x)\geq 0$, which completes the proof.\\

Let us define $\bar{h}_\epsilon (t,y,A)= - h_\epsilon (t,y)\chi _A(y) +
\int_A h_\epsilon (t,y,x)dx$. It is obvious that $\bar{h}_\epsilon $ is a charge (that
is a real-valued measure  with the property $\bar{h}_\epsilon (t,y,R)=0$),
\cite{gihman}.

We shall show that there exists a \it step \rm process corresponding to the
charge $\bar{h}_\epsilon $.

To this end let us first prove: \\

{\bf Lemma 4}\\

For any Borel  set $A\subset R$, the function
$t\rightarrow \int_A h_\epsilon (t,y,x)dx$ 
is continuous in $t$, uniformly in $A$.  \\

{\bf Proof}:\\

We have the following estimate (cf. Eq. (18) and Lemma 3):
$$|\int_A h_\epsilon (t,y,x)dx - \int_Ah_\epsilon (t_0,y,x)dx | =
|\int_A{{\theta ^\epsilon (y+x,t)}\over {\theta ^\epsilon (y,t)}}
q_\epsilon (x)dx   - \int_A {{\theta ^\epsilon (y+x,t_0)}\over
{\theta ^\epsilon (y,t_0)}} q_\epsilon (x) dx|\leq $$
$$| \int_{A\cap K^c} [{{\theta ^\epsilon (y+x,t)}
\over {\theta ^\epsilon (y,t)}} - {{\theta ^\epsilon (y+x,t_0)}
\over {\theta ^\epsilon (y,t)}}] q_\epsilon (x)dx| + |\int_{A\cap K}
[{{\theta ^\epsilon (y+x,t)}\over {\theta ^\epsilon (y,t)}} -
{{\theta ^\epsilon (y+x,t_0)}\over {\theta ^\epsilon (y,t)}}]
q_\epsilon (x) dx| + $$
$$|\int_A  [{{\theta ^\epsilon (y+x,t_0)}\over {\theta ^\epsilon (y,t)}} -
{{\theta ^\epsilon (y+x,t_0)}\over {\theta ^\epsilon (y,t_0)}}]
q_\epsilon (x) dx |$$
where $K$ is a compact set while  $K^c$ is its complement. \\
Let us denote the summands $A_1, A_2, A_3$ respectively.
For the first summand we have
$$A_1 \leq  {1\over {\theta ^\epsilon (y,t)}}\,  sup_{x\in R} \,
(\theta ^\epsilon (x,t) +  \theta ^\epsilon (x,t_0))\,
\int_{K^c} q_\epsilon (x) dx\, . $$
But:
$$sup_{x\in R}\, \theta ^\epsilon (x,t) = sup_{x\in R} \int_R k_\epsilon
(T-t,x-y)
g(y)dy \leq M\, sup_{x\in R} \int_R k_\epsilon (T-t,x-y)dy = M $$
By defining $N(y)= sup_{t\in [t_0,t_0+ 1]}
{1\over {\theta ^\epsilon (y,t)}}$ and adjusting the compact set $K$ so that
$\int_{K^c} q_\epsilon (x)dx \leq {\delta \over {3MN(y)}}$, we arrive at
$A_1 \leq {\delta \over 3}$.\\
With  the second summand,  $A_2$,  we proceed as follows:
$$A_2 = |\int_{A\cap K} [{{\theta ^\epsilon (x,t)}\over
{\theta ^\epsilon  (y,t)}} -
{{\theta ^\epsilon (x,t_0)}\over {\theta ^\epsilon (y,t)}}]
q_\epsilon (y-x) dx |\leq
 N(y) \, sup_{x\in K} \, |\theta ^\epsilon (x,t) - \theta ^\epsilon (x,t_0)|
{2\over {\pi \epsilon }}$$
By choosing $t$ so close to $t_0$ that $sup_{x\in K} |\theta ^\epsilon (x,t) -
\theta ^\epsilon (x,t_0)|\leq {{\pi \delta \epsilon }\over {6N(y)}}$,
we get $A_2 \leq {\delta \over 3}$.\\
Analogously with $A_3$:
$$A_3 \leq |{1\over {\theta ^\epsilon (y,t)}} - {1\over
{\theta ^\epsilon (y,t_0)}}|\,
 2\, sup_{x\in R} \theta ^\epsilon (x,t_0) {2 \over {\pi \epsilon }}
 \leq
 {4\over {\pi  \epsilon }} M N^2(y) |\theta ^\epsilon (y,t_0) -
 \theta ^\epsilon (y,t)|$$
 where by taking $t$ such that $|\theta ^\epsilon (y,t_0) -
 \theta ^\epsilon (y,t)|\leq
 {{\pi \delta \epsilon }\over {12MN^2(y)}}$ we shall get $A_3
 \leq {\delta \over 3}$.
 The overall bound is thus $\delta $, and the Lemma is proved.\\

As a byproduct of the  above demonstration, we realise  that the
function $t\rightarrow h_\epsilon (t,x,A)$   is continuous in
$t$ uniformly on compact sets.
As a consequence, see e.g. Theorem 4 in chap. 7, sec. 7 of Ref.
\cite{gihman}, there exists a stochastically
continuous  Markov process
$X^\epsilon _t$ with continuous from the right sample paths.
Moreover, for any
$s\in [0,T]$, $y\in R$ and $A\subset R$, there holds:
$${lim_{t\downarrow s} {{p_\epsilon (y,s,A,t) - \chi _A(y)}\over {t-s}} =
\bar{h}_\epsilon (s,y,A)}\eqno (19)$$
where $p_\epsilon (y,s,A,t)$ is the transition kernel of the process
$X^{\epsilon }_t$. \\
There follows: \\

{\bf Theorem 3} \\

The transition probability density of $X^\epsilon _t$ reads:
$$p_\epsilon (y,s,x,t)= k_\epsilon (t-s,x-y){{\theta ^\epsilon (x,t)}
\over {\theta ^\epsilon (y,s)}}$$
and   is a solution of  the first Kolmogorov equation:
$$\partial _sp_\epsilon (y,s,x,t) = - \int_R p_\epsilon (z,s,x,t)
\bar{h}_\epsilon (s,y,z)dz$$

{\bf Proof}:\\

We must demonstrate that Eq. (19) is valid for the just introduced
transition density  (compare e.g. also Theorem 1), i.e. there holds:
$$lim_{t\downarrow s} {1\over {t-s}} [k_\epsilon (t-s,x-y)
{{\theta ^\epsilon (x,t)}\over
{\theta ^\epsilon (y,s)}} - \delta _y(x)]= \bar{h}_\epsilon (s,y,x)\, .$$
To this end, let us notice (adding and subtracting the same summand) that
$$\bar{h}_\epsilon (s,y,x) = $$
$${{\theta ^\epsilon (x,s)}\over {\theta ^\epsilon (y,s)}} lim_{t\downarrow s}
{1\over {t-s}} [k_\epsilon (t-s,x-y) - \delta _y(x)]  +
{{\delta _y(x)}\over {\theta ^\epsilon (y,s)}} lim_{t\downarrow s}
{1\over {t-s}}[\theta ^\epsilon (x,t) - \theta ^\epsilon (y,s)] = $$
$${{\theta ^\epsilon (x,s)}\over {\theta ^\epsilon (y,s)}} [q_\epsilon (x-y) -
{2\over {\pi \epsilon }} \delta _y(x)] + {{\delta _y(x)}\over {\theta ^\epsilon (y,s)}}
lim_{t\downarrow s} {1\over {t-s}} [\theta ^\epsilon (x,t) - \theta ^\epsilon (y,s)]\, .
$$
To evaluate  the second term, let us take a continuous and bounded
function $a(x)$  and consider
$$lim_{t\downarrow s} \int_R {{\delta _y(x)}\over {\theta _\epsilon (y,s)}}
{1\over {t-s}} [\theta ^\epsilon (x,t) - \theta ^\epsilon (y,s)] a(x) dx = $$
$$lim_{t\downarrow s} {{a(y)}\over {\theta ^\epsilon (y,s)}} {1\over {t-s}}
[\theta ^\epsilon (y,t) -
\theta ^\epsilon (y,s)] = {{a(y)}\over {\theta ^\epsilon (y,s)}}
\partial _s \theta ^\epsilon (y,s)\, .$$
So,  the second term converges weakly  to
$${{\delta _y(x)}\over {\theta ^\epsilon (y,s)}}
\partial _s \theta ^\epsilon (y,s)\, .$$
We know that
$$\partial _s\theta ^\epsilon (y,s)= |\nabla |_\epsilon
\theta ^\epsilon (y,s) =
- \int_R [\theta ^\epsilon (y+z,s)-\theta ^\epsilon (y,s)]
q_\epsilon (z) dz\, .$$
Consequently
$${{\partial _s \theta ^\epsilon (y,s)}\over
{\theta ^\epsilon (y,s)}} = - \int_R
{{\theta ^\epsilon (y+z,s)}\over {\theta ^\epsilon  (y,s)}}
q_\epsilon (z)dz +
{2\over {\pi \epsilon }} =
{2\over {\pi \epsilon }} - h_\epsilon (s,y)$$
and thus
$$lim_{t\downarrow s} {1\over {t-s}} [p_\epsilon (y,s,x,t) -
\delta _y(x)] = $$
$$
{{\theta ^\epsilon (x,s)}\over {\theta ^\epsilon (y,s)}}
q_\epsilon (x-y) - {2\over {\pi \epsilon }}
\delta _y(x) + {2\over {\pi \epsilon }}\delta _y(x)  -
h_\epsilon (s,y) \delta _y(x) =$$
$$ h_\epsilon (s,y,x) - h_\epsilon (s,y) \delta _y(x) =
\bar{h}_\epsilon (s,y,x)$$
The first part of our Theorem  is proved, and we can pass to its second part.

To check the validity of the Kolmogorov equation, we shall begin from
$$\partial _sp_\epsilon (y,s,x,t) = [\partial _s k_\epsilon (t-s,x-y)]
{{\theta ^\epsilon (x,t)}\over {\theta _\epsilon (y,s)}} -
p_\epsilon (y,s,x,t) {{ \partial _s\theta ^\epsilon (y,s)}
\over {\theta ^\epsilon (y,s)}}$$
But:
$$\partial _sk_\epsilon (t-s,x-y) = - [q_\epsilon *k_\epsilon (t-s,.)](x-y) +
k_\epsilon (x-y) {2\over {\pi \epsilon }}$$
and
$${{\partial _s \theta ^\epsilon (y,s)}\over {\theta _\epsilon (y,s)}} =
{2\over {\pi \epsilon }} - h_\epsilon (s,y)$$
which leads to
$$\partial _sp(y,s,x,t)=$$
$$ - [q_\epsilon *k_\epsilon (t-s,.)](x-y)
{{\theta ^\epsilon (x,t)}\over {\theta ^\epsilon (y,s)}} +
{2\over {\pi \epsilon }}p_\epsilon (y,s,x,t) -$$
$${2\over {\pi \epsilon }}p_\epsilon (y,s,x,t) + p_\epsilon (y,s,x,t)
h_\epsilon (s,y) =$$
$$- {{\theta ^\epsilon (x,t)}\over {\theta ^\epsilon (y,s)}}
\int_R q_\epsilon (x-y-z)
k_\epsilon (t-s,z) dz + p_\epsilon (y,s,x,t) \int_R
{{\theta ^\epsilon (x+y,s)}\over
{\theta ^\epsilon (y,s)}} q_\epsilon (x) dx \, .$$
On the other hand
$$- \int_R p_\epsilon (z,s,x,t) \bar{h}(s,y,z)dz = $$
$$- \int_R k_\epsilon (t-s,x-z)
{{\theta ^\epsilon  (x,t)}\over {\theta ^\epsilon (z,s)}}
\, [{{\theta ^\epsilon (z,s)}\over {\theta ^\epsilon (y,s)}}q_\epsilon (z-y) -
\delta _y(z) h_\epsilon (s,y)]dz = $$
$$- {{\theta ^\epsilon (x,t)}\over {\theta _\epsilon (y,s)}}\int_R
k_\epsilon (t-s,x-z)q_\epsilon (z-y) dz +
p_\epsilon (y,s,x,t)h_\epsilon (s,y)\, . $$
Since  we know that $h_\epsilon (s,y) = \int_R {{\theta ^\epsilon (x+y,s)}
\over {\theta _\epsilon (y,s)}} q_\epsilon (x) dx $,  the assertion
(e.g. the validity of the
first Kolmogorov equation) follows.\\

{\bf Corollary}\\

$X^\epsilon _t$ is a step process. \\

{\bf Proof}:\\

It suffices to check  that $p_\epsilon (y,s,R,t)=1$  (cf.
Ref. \cite{gihman}).
Since
$$p_\epsilon (y,s,R,t)= \int_R p_\epsilon (y,s,x,t)dx =
\int_R k_\epsilon (t-s,x-y){{\theta ^\epsilon (x,t)}\over
{\theta ^\epsilon (y,s)}} dx $$
and, by Theorem 2,
$$\int_R k_\epsilon (t-s,x-y)\theta ^\epsilon (x,t)dx =
\theta ^\epsilon (y,s)$$
the    Corollary holds true.\\

All previous considerations  can be finally summarized
by  showing that  the  family
$X^\epsilon _t$  of  step processes  consistently  approximates
(converges to) the process $X_t$.
Indeed, we have:   \\

{\bf Theorem 4}\\

The limit:            \\
$$lim_{\epsilon \downarrow 0} X^\epsilon _t = X_t$$    \\
holds true in distributions  and uniformly in $t\in [0,T]$.
Moreover,  the transition probability density $p_\epsilon $
converges pointwise to $p$ when $\epsilon \downarrow 0$.

{\bf Proof}:\\

The probability density of the process $X^\epsilon _t$  equals
to $\rho _\epsilon (x,t) = \theta ^\epsilon _*(x,t) \theta ^\epsilon (x,t)$
and      that of the process $X_t$ is given by $\rho (x,t) =
\theta _*(x,t) \theta (x,t)$. But,   $\theta _*^\epsilon (x,t) =
\int_R k_\epsilon (t,x-y) f(y) dy$  and  $k_\epsilon (t,x-y)$
converges weakly to the
Cauchy kernel $k(t,x-y)$, uniformly in $t$. Consequently
$lim_{\epsilon \downarrow 0}\theta _*^\epsilon (x,t) = \theta _*(x,t)$
also uniformly in $t\in [0,T]$.
The same holds true for $\theta ^\epsilon (x,t)$, and the
first assertion follows.

The second statement follows from the fact that $k_\epsilon (t,x) $ tends
to  the Cauchy kernel $k(t,x)$ (see Lemma 2) when $\epsilon \downarrow 0$.\\

As stated before,  considerations of the present section were mostly a
preparation to the study of perturbed problems.  However,
it is useful to mention
that the conditional Cauchy processes are covered by the developed scheme.
In fact, we can here adjust to the Cauchy noise
an observation previously
utilized in the context of the Wiener noise, \cite{jam,olk,klaud}.
 The pertinent
density can be given in the following form:
$${\rho (x,t) = {{k(y_0,t_0,x,t) k(x,t,z_T,T)}\over
{k(y_0,t_0,z_T,T)}}}\eqno (20) $$
with $y_0,z_T \in R$ and $0<t_0<t<T$. All previous considerations directly apply
 to the interpolating process supported by this density. See also for a discussion of
 L\'{e}vy bridges (while specialised to the Cauchy context) in
 Ref. \cite{bertoin}.

\section{Perturbations of the Cauchy noise}

An important conceptual input in probabilistic solutions of the
Schr\"{o}dinger   interpolation problem was  the clean
identification   of the r\^{o}le played by  the Feynman-Kac kernels,
specifically by their joint continuity in spatial variables.
This technical feature received proper attention in constructions based on the
conditional Wiener measure, \cite{sim,simon}, but no analogous results seem
to be in existence relative to other conditional measures, even if the
pertinent  process and its sample paths are deduced from  an
infinitely divisible probability law (this issue we have analyzed in
the previous section).
The same obstacle appears in the context
of perturbed processes, where the Feynman-Kac formula is known to be valid,
\cite{carm,carm1,jacob}, but the relevant  properties of the Feynman-Kac
kernels  have  not been  investigated in the literature.
\\
We are motivated by the strategy of Refs. \cite{olk,klaud},
and the techniques developed in the previous section .
Let us address the problem analogous to that of Eq. (11), but
now in reference to a perturbed semigroup, \cite{carm}:
$${\partial _t \theta _* = - |\nabla | \theta _* - V\theta _*}\eqno (21)$$
$$\partial _t \theta = |\nabla |\theta + V\theta $$
where $V$ is a measurable function such that:\\
(a) for all $x\in R$, $V(x)\geq 0$, \\
(b) for each  compact set $K\subset R$  there exists $C_K$ such that
for all $x\in K$, $V$ is locally bounded $V(x)\leq C_K$.\\
Then $V$ is locally integrable and for any compact $K$  we have
$${lim_{t\downarrow 0}\, sup_{x\in R} \, E^C_x\{ \int_0^t \chi _K(X^C_s)
V(X^C_s) ds\} = 0\, .}\eqno (22)$$
As a consequence, there holds\\

{\bf Lemma 5} \\

If $1\leq r\leq p\leq \infty $ and $t>0$, then the operators $T^V_t$
defined by
$$(T^V_tf)(x) = E^C_x\{ f(X^C_t) exp[-\int_0^t V(X^C_s)ds]\}$$
are bounded from $L^r(R)$ into $L^p(R)$. Moreover, for each
$r\in [1,\infty ]$ and $f\in L^r(R)$, $T_t^Vf$ is a bounded and
continuous function. \\

{\bf Proof}:\\

See e.g. Ref. \cite{carm}, Proposition III.1.\\

We shall also use  another identity proved by Carmona, \cite{carm},
namely:\\

{\bf Lemma 6}\\

For any real-valued  $f,g \in L^2(R)$  there holds
$$\int_R dx\, f(x) E^C_x\{ g(X^C_t) exp[ - \int_0^t V(X_s^C)ds]\} =  $$
$$
\int_R dx \, g(x) E_x^C\{f(X^C_t) exp[-\int_0^t V(X^C_s)ds]\}\, .$$

{\bf Proof}:\\

Cf. Eq. (III.9)  in Ref. \cite{carm}. \\

We need to prove that $T^V_t$ is an integral operator.
To this end,
a direct  transfer of Simon's arguments, cf.
Ref. \cite{simon1}, originally with respect to the Laplace
differential operator,
i. e. the usage   of the Dunford-Pettis theorem
(see pp. 450 in \cite{simon1})  and Lemma 5, gives rise to: \\

{\bf Lemma 7}\\

For any $p\in [1,\infty ]$ and $f\in L^p(R)$ there holds
$$(T_t^Vf)(x) = \int_R k^V_t (x,y) f(y) dy $$
where $k^V_t(x,y) \geq 0$ almost everywhere and, for $q$ such that
${1\over q} + {1\over p}=1$, the kernel satisfies
$$sup_{x\in R} [\int_R [k^V_t(x,y)]^qdy ]^{1/q} < \infty $$

{\bf Proof}: \\

Cf. Theorem A.1.1 and Corollary A.1.2 in Ref. \cite{simon1}.\\

Notice that by putting $p=1$ and thus $q=\infty $ we  obtain that
$k^V_t(x,y) \in L^{\infty }(R^2)$.

Our ultimate goal is to utilize  $k^V_t(x,y)$ in the context of the
Schr\"{o}dinger boundary data and interpolation problem, \cite{jam,olk},
hence suitable properties of the kernel must be established.
For our purposes, the joint continuity and positivity of the kernel
is essential.\\

{\bf Lemma 8}\\

$k^V_t(x,y)$ is jointly continuous in  $(x,y)$.\\

{\bf  Proof}:\\

We begin from demonstrating that $k^V_t(x,y) = k^V_t(y,x)$ almost
everywhere.   \\
By  Lemma 6, we have
$$\int \int_{R^2} dx\, dy\, f(x) k^V_t(x,y) g(y) =
\int \int_{R^2} dx\, dy\, g(x)k^V_t(x,y) f(y) \, ,$$
hence
$$\int \int_{R^2} dx\, dy\, f(x)g(y) [k^V_t(x,y) - k^V_t(y,x)] =0$$
for all $f,g\in L^2(R)\cap L^1(R)$. \\
The same holds true  for all finite combinations
$\Sigma _{i,j} a_{ij} f_i(x)g_j(y)$. Therefore
$\int \int_{R^2} [k^V_t(x,y) - k^V_t(y,x)]f(x,y)dx\, dy = 0 $
for all $f(x,y)$ from  a  dense subset of $L^1(R^2)$.
Because $L^{\infty }(R^2)$ is the dual space to $L^1(R^2)$, we conclude
that $k^V_t(x,y)=k^V_t(y,x)$ almost everywhere.\\
Let us exploit the semigroup property of $k^V_t(x,y)$:
$$k^V_t(x,y)= \int_R k^V_{t/2}(x,w) k^V_{t/2}(w,y) dw\, .$$
For each  $y$, $w\rightarrow k^V_{t/2}(w,y) \in L^{\infty }(R)$
so, by Lemma  5,  $k^V_t(x,y)$ is continuous in $x$. By the symmetry,
$k^V_t(x,y)$ is separately continuous in $x$ and $y$. \\
Let us consider a sequence $(x_n,y_n)\rightarrow (x,y)$. Then:
$$|k^V_t(x_n,y_n) - k^V_t(x_0,y_0)| \leq $$
$$ |\int \int_{R^2} dw dz [k^V_{t/3}(x_n,w) - k_{t/3}^V(x_0,w)] k^V_{t/3}(w,z)
k^V_{t/3}(z,y_n)| + $$
$$|\int \int_{R^2} dw dz k^V_{t/3}(x_0,w)k^V_{t/3}(w,z) [k^V_{t/3}(z,y_n) -
k^V_{t/3}(z,y_))]|=$$
$$|\int_R dw [k^V_{t/3}(x_n,w) - k^V_{t/3}(x_0,w)]k^V_{2t/3}(w,y_n)| +
|k^V_t(x_0,y_n)-k^V_t(x_0,y_0)|\, .$$
Because of
$$||k^V_{2t/3}(.,y_n)||_{L^\infty }  <  C $$
for all $y_n$, knowing that $sup_{n} k^V_{t/3}(x_n,w)$ exists and is
integrable     with respect to $w$, by the Lebesgue dominated convergence
theorem   the first summand tends to zero. \\
Hence, $k^V_t(x,y)$ is jointly continuous in $(x,y)$.\\

{\bf Lemma 9}\\

$k^V_t(x,y)$ is strictly positive.\\

{\bf Proof}:\\

Because for the Cauchy process we have, \cite{weron} (more general
estimates of the growth of
random walks  and L\'{e}vy processes can be found in \cite{pruitt}):
$$E^C_x\{sup_{0\leq s\leq t}\,  |X^C_s| > n \} \leq
3 sup_{0\leq s\leq t}\,  E^C_x\{ |X^C_s|>{n\over 3}\}$$
and
$$sup_{0\leq s\leq t}\,  E^C_x\{ |X^C_s| > {n\over 3}\} =
E_x^C\{ |X^C_t|>{n\over 3}\} =
1 - {2\over \pi } arctan \, ({n\over {3t}})$$
there follows:
$$lim_{n\rightarrow \infty } E^C_x\{ sup_{0\leq s\leq t} |X^C_s|> n\}= 0$$
This property will be used  in below.\\
Let $0< \delta < 1$, then:
$$\int_{y-\delta }^{y+\delta } dy k^V_t(x,y) = E^C_x\{
\chi _{[y-\delta ,y+\delta ]}(X^C_t) \, exp [- \int_0^t V(X^C_s) ds]\}\, . $$
By the previously deduced property, for fixed $x$ and $y$,
 we can choose a compact set $[-n,n]$ such that
 $$E^C_x\{ \Omega _{(t,[y-\delta,y+\delta ])}^{(0,x)} (n) \} > {1\over 2}
 \int_{y-\delta }^{y+\delta }k_t(x,y) dy$$
 where
 $$\Omega ^{(0,x)}_{(t,[y-\delta ,y+\delta ])} (n) = \{ \omega :
 \omega (0)=x, \omega (t) \in [y-\delta, y+\delta ];  s\in [0,t]
 \Rightarrow \omega (s)\in [-n,n]\}$$
 and $k_t(x,y)$ is the Cauchy kernel.
 Hence
 $$\int_{y-\delta }^{y+\delta }dy k^V_t(x,y) \geq \int_{\Omega  (n)}
 exp[-\int_0^t V(X^C_s)ds]dP_x^C(\omega ) \geq {1\over 2} exp(-c_nt)\,
 \cdot \int_{y-\delta }^{y+\delta } k_t(x,y)dy $$
 where $c_n = sup_{x\in [-n,n]}V(x)$.

 Because $k^V_t(x,y)$ is continuous and $\delta $ was arbitrary, we get
 $$k^V_t(x,y) \geq {1\over 2} exp(-c_nt)\, \cdot k_t(x,y)$$
 The assertion  of  Lemma 9  is thus valid. \\

Lemma 8 and 9  provide us with a strictly positive and jointly
continuous in space variables kernel, which can be directly exploited
for the analysis of the Schr\"{o}dinger interpolation problem,
as exemplified by Eqs. (3)- (6), see also \cite{jam,zambr,olk}.
Indeed, let $\rho _0(x)$ and $\rho _T(x)$  be strictly positive densities.
 Then, the Markov process $X^V_t$ characterized by the transition
 probability density:
 $${p^V(y,s,x,t)= k^V_{t-s} (x,y) {{\theta (x,t)}\over
 {\theta (y,s)}} }\eqno (21)$$
and the density of distributions
$${\rho (x,t)= \theta _*(x,t)\theta (x,t)}\eqno (22)$$
where:
$$\theta _*(x,t) = \int_R k^V_t(x,y)f(y)dy$$
$${\theta _*(y,t) = \int_R k^V_{T-t}(x,y)g(x)dx}\eqno (23)$$
is precisely that  interpolating Markov process to which
Theorem 1 extends its validity,  when the perturbed semigroup kernel
replaces the Cauchy kernel.

Clearly,  for all $0\leq s\leq t\leq T$ we have
$$\theta _*(x,t)=\int_R k^V_{t-s}(x,y) \theta _*(y,s) dy$$
$${\theta (y,s) = \int_Rk^V_{t-s}(x,y)\theta (x,t)dx}\eqno (24)$$
and that suffices for the Theorem 1 to hold true in the present
case as well.\\

Following the strategy of the previous section, we shall investigate
an issue of approximating the perturbed Cauchy process
(set by Lemmas 8, 9 and Theorem 1)  by means of step processes.

Let us first invoke the step process $Y^\epsilon _t$ of Lemma 1.
It corresponds to the unperturbed generator $|\nabla |_\epsilon $.
To account for a perturbation and the involved perturbed semigroup,
  let us consider a multiplicative, homogeneous and
contracting   functional:
$${\alpha ^s_t(\omega ) = exp[-\int _s^t
V(Y^\epsilon _\tau (\omega ))d\tau ]}\eqno (25)$$
of the process $Y^\epsilon _t$,  for times
$0\leq s\leq t\leq T$.

  We recall that the process $Y^\epsilon _t$
is a step process    obtained from  the Cauchy process by
neglecting "small jumps" (the $\epsilon$-cutoff).

We shall associate with the multiplicative functional (25) the process
$Y^{\epsilon,V}_t$ and prove that   under additional restrictions on the
potential   $V$, the pertinent perturbed process is also a step process.  \\

{\bf Theorem 5}\\

Let $0\leq V(x)\leq M$ for all $x\in R$.
The transition function:
$$p_{\epsilon,V}(t,x,\Gamma )= E^\epsilon _x\{ \chi _\Gamma (Y^\epsilon _t)
exp[-\int_0^tV(Y_s^\epsilon )ds] \}$$
determines the step process $Y^{\epsilon, V}_t$.\\

{\bf Proof}:\\

By Theorem 3.8 of Ref. \cite{dynkin}  a sufficient condition for
the existence of a Markovian step process $Y^{\epsilon ,V}_t$
is that  its transition function obeys
$$lim_{t\downarrow 0}\,  p_{\epsilon,V}(t,x,\{ x\})= 1$$
uniformly in $x\in R$.

Let us choose $t_1> 0$  so that $1-\delta \leq exp(-Mt_1)$ is secured.
In view of
$$exp(-Mt) \leq exp[-\int_0^tV(Y^\epsilon _s(\omega ))ds]  \leq 1$$
for all $\omega $, we have for all $t<t_1$ the following estimate:
$$(1-\delta )p_\epsilon (t,x,\Gamma ) \leq p_{\epsilon, V}(t,x,\Gamma )
\leq p_\epsilon (t,x,\Gamma )\, .$$
On the other hand, there exists $t_2$ such that  for all $t<t_2$
$$p_\epsilon (t,x,\{ x\}) \geq 1-\delta $$
is valid for all $x\in R$. \\
Hence, for all $t<min(t_1,t_2)$ we get
$$(1-\delta )^2 \leq p_{\epsilon ,V}(t,x,\{ x\} )\leq 1$$
Because $\delta $ is arbitrary, after taking $\delta \rightarrow 0$,
the assertion follows.\\

From the formula $p_{\epsilon ,V}(t,x,\Gamma ) \leq
p_{\epsilon } (t,x,\Gamma )$  we conclude that the transition function
$p_{\epsilon , V}(t,x,\Gamma )$ is absolutely continuous with respect to the
Lebesgue measure, and  hence posesses  a density $k_{\epsilon ,V}(t,x,y)$.

A new process $X^{\epsilon ,V}_t$ can be defined by considering a
multiplicative transformation  of the process $Y_t^{\epsilon ,V}$
by means of
$${\alpha _s^t = {{\theta ^\epsilon (Y_t^{\epsilon ,V}, t)}\over
{\theta ^\epsilon  (Y_s^{\epsilon ,V},s)}}\, ,}\eqno (26) $$
where $\theta ^{\epsilon }$ is a solution of $\partial _t
\theta ^{\epsilon } = |\nabla |_{\epsilon } \theta ^{\epsilon } +
V\theta ^{\epsilon }$.

The transition probability density of $X_t^{\epsilon ,V}$ reads
$${p_{\epsilon ,V} (s,y,t,x) = k_{\epsilon ,V}(t-s,y,x)
{{\theta ^\epsilon (x,t)}
\over {\theta ^\epsilon (y,s)}}}\eqno (27)$$
and by repeating  arguments mimicking  those of Section II, one
can show that
the perturbed step process  $X_t^{\epsilon ,V}$ converges in
distribution to the  perturbed Cauchy  process $X^V_t$, when
$\epsilon \rightarrow 0$, uniformly in $t\in [0,T]$.
\\

A concise summary  of all  mathematical arguments of sections II
and III, reads:   \\
(a) We have found a solution of the Schr\"{o}dinger interpolation problem
whose kernel function is determined by the Cauchy generator plus
a potential.\\
(b) We have described the pertinent process (and its simpler versions,like
the conditional  Cauchy process of section II) as a limit of \it step \rm
processes. \\
(c) The devoloped  techniques can be used to investigate  the
existence issue  (including that of  the step process approximation)
of more general jump-type processes, in particular those  related to the
quantum evolution with relativistic Hamiltonians,
\cite{klaud,deang}.\\

{\bf Remark}:\\
In the present paper, to simplify calculations and to make formulas more
transparent, we have considered processes associated with the Cauchy generator
(and thus with the $\alpha $- stable symmetric process as a major tool)
in space dimension $1$.  A glance at the construction of solutions
of the Schr\"{o}dinger problem  makes clear that the previous
limitations are inessential.  In fact, we could consider any
$\alpha \in (0,2)$ - symmetric stable processes on $R^n$, for
 arbitrary $n \geq 1$, and secure  the strict positivity and joint continuity
 in space variables of the corresponding transition density.
 Such properies for $n \geq 2$ and for potentials  from the Kato class
 $K_{n, \alpha }$   were established  in the very recent publication,
 \cite{chen}, Theorems 3. 3 and 3. 5.

{\bf Acknowledgements}: \\
 Both authors are willing words of gratitude to
Professor Ph. Blanchard, whose hospitality at the
Bielefeld-Bonn-Stochastic Research  Center at the University of Bielefeld
made this collaboration possible. R. O. would like to thank the
Alexander von Humboldt  Foundation for financial support.
P. G. would like to thank Professor R. Carmona and Professor
W. Woyczynski for   correspondence on the issue
of Feynman-Kac kernels in non-Wiener contexts.


\begin{thebibliography}{99}
\bibitem{schr} E. Schr\"{o}dinger, Ann. Inst. Henri Poincar\'{e}, 
{\bf  2}, 269, (1932)

\bibitem{jam}
 B. Jamison,  Z. Wahrsch.  verw. Geb. {\bf 30}, 65, (1974) 


\bibitem{zambr}  J. C. Zambrini, 
 J. Math. Phys.  {\bf 27}, 2307, (1986)

\bibitem{garb} J. C. Zambrini, pp. 393 in: {\it Chaos-The Interplay Between Stochastic and
Deterministic Behaviour},
Karpacz'95 Proc., LNP vol. 457, P. Garbaczewski, M. Wolf,
A. Weron (eds.),  (Springer-Verlag, Berlin 1995).

\bibitem{klaud} P.Garbaczewski, J. R. Klauder, R. Olkiewicz,
{\it Phys. Rev.} {\bf E 51}, 4114 (1995).



\bibitem{olk}  P. Garbaczewski, R. Olkiewicz, 
 J. Math. Phys.  {\bf 37}, 731, (1996)

\bibitem{kondrat} P. Garbaczewski, G. Kondrat, Phys. Rev. Lett. {\bf 77},
2608, (1996)

\bibitem{blanch}  Ph. Blanchard, P. Garbaczewski, R. Olkiewicz,
J. Math. Phys.  {\bf 38}, 1, (1997)


\bibitem{sim} M. Reed, B.Simon,
{\it Methods of Modern Mathematical Physics}, vol. IV,
(Academic, New York 1978)

\bibitem{simon}  B. Simon, {\it Functional Integration and Quantum Physics},
(Academic, New York, 1979)

\bibitem{carm}  R. Carmona, pp. 65 in: {\it Schr\"{o}dinger Operators}, LNP vol.
345, H. Holden, A. Jensen (eds.), (Springer-Verlag, Berlin, 1989)

\bibitem{carm1} R. Carmona, W. C. Masters, B. Simon, J. Funct. Anal. {\bf 91},
117, (1990)

\bibitem{jacob} N. Jacob, {\it Pseudo-Differential   Operators and Markov
Processes},  (Akademie Verlag, Berlin, 1996)

\bibitem{shl}  M. F. Shlesinger, G. M. Zaslavsky, U. Frisch, (eds.) {\it
L\'{e}vy Flights and Related Topics in Physics}, LNP vol. 450,
(Springer-Veerlag, Berlin, 1995)

\bibitem{west} B. J. West, V. Seshadri, Physica {\bf 113 A}, 203, (1982)

\bibitem{monti} F. Monti, H. R. Jauslin, J. Stat. Phys. {\bf 60}, 413,
(1990)

\bibitem{tsallis} C. Tsallis et al. Phys. Rev. Lett. {\bf 75}, 3589, (1995)

\bibitem{grigo} G. Tref\'{a}n et al. Phys. Rev. {\bf E 50}, 2564, (1994)


\bibitem{dynkin} E. B. Dynkin, {\it Markov Processes}, vol. I, (Springer-Verlag,
Berlin, 1965)

\bibitem{gihman} I. I. Gihman, A. V. Skorohod, {\it Introduction to
the Theory of Random Processes},  (W. B. Saunders Comp., Philadelphia,
1969)

\bibitem{brei} L. Breiman, {\it Probability}, (Addison-Wesley, Reading, 1968)

\bibitem{bil} P. Billingsley, {\it Probability and Measure}, (Wiley, New York,
1979)

\bibitem{jacod} J. Jacod, A. N. Shiryaev, {\it Limit Theorems for Stochastic
Processes}, (Springer-Verlag, Berlin, 1987)

\bibitem{kurtz} S. N. Ethier, T. G. Kurtz, {\it Markov Processes:
Characterization  and Convergence}, (Wiley, New York, 1986)

\bibitem{bertoin} J. Bertoin, {\it L\'{e}vy Processes}, (Cambridge
University Press, Cambridge, 1996)

\bibitem{simon1}  B. Simon, Bull. Amer. Math. Soc. {\bf 7}, 447, (1982)

\bibitem{weron}  A. Janicki, A. Weron, {\it Simulation and Chaotic
Behaviour of $\alpha$-stable Stochastic Processes},
(M. Dekker, New York, 1994)

\bibitem{pruitt}  W. E. Pruitt,   Ann. Prob. {\bf 9},  948, (1981)

\bibitem{deang}    G. F. De Angelis, J. Math. Phys. {\bf 31}, 1408, (1990)

\bibitem{chen} Z. Q. Chen, R. Song, J. Funct. Anal. {\bf 150}, 204, (1997)


\end{thebibliography}
\end{document}